\documentclass[aps,prl,amsmath,amssymb,amstext,citeautoscript,punctuation,nofootinbib,superscriptaddress,twocolumn]{revtex4-1}
\usepackage[dvips]{graphicx}
\usepackage{amsmath,amssymb,amsthm,mathrsfs,amsfonts,dsfont,hyperref,subfigure, epsfig, braket,bm,enumerate,color,graphicx,dcolumn}
\usepackage[normalem]{ulem}
\usepackage[dvipsnames]{xcolor}
\usepackage{{xcolor,charter, gensymb, float}}
\usepackage [english]{babel}

\newcommand{\eca}{EuCuAs}
\begin{document}
\title{Weyl metallic state induced by helical magnetic order}
\author{Jian-Rui Soh}
\affiliation{Institute of Physics, Ecole Polytechnique Fédérale de Lausanne (EPFL), CH-1015 Lausanne, Switzerland}
\author{Irián Sánchez-Ramírez}
\affiliation{Donostia International Physics Center, P. Manuel de Lardizabal 4, 20018 Donostia-San, Sebastian, Spain}
\author{Xupeng Yang}
\affiliation{Institute of Physics, Ecole Polytechnique Fédérale de Lausanne (EPFL), CH-1015 Lausanne, Switzerland}
\author{Jinzhao Sun}
\affiliation{Department of Physics, University of Oxford, Clarendon Laboratory, Oxford OX1 3PU, United Kingdom}
\author{Ivica Zivkovic}
\affiliation{Institute of Physics, Ecole Polytechnique Fédérale de Lausanne (EPFL), CH-1015 Lausanne, Switzerland}
\author{J. Alberto Rodr\'iguez-Velamaz\'an}
\affiliation{Institut Laue-Langevin, 6 rue Jules Horowitz, BP 156, 38042 Grenoble Cedex 9, France}
\author{Oscar Fabelo}
\affiliation{Institut Laue-Langevin, 6 rue Jules Horowitz, BP 156, 38042 Grenoble Cedex 9, France}
\author{Anne Stunault}
\affiliation{Institut Laue-Langevin, 6 rue Jules Horowitz, BP 156, 38042 Grenoble Cedex 9, France}
\author{Alessandro Bombardi}
\affiliation{Diamond Light Source, OX11 0DE, United Kingdom}
\author{Christian Balz}
\affiliation{ISIS, Rutherford Appleton Laboratory - STFC, OX11 0QX, United Kingdom}
\author{Manh Duc Le}
\affiliation{ISIS, Rutherford Appleton Laboratory - STFC, OX11 0QX, United Kingdom}
\author{Helen C. Walker}
\affiliation{ISIS, Rutherford Appleton Laboratory - STFC, OX11 0QX, United Kingdom}
\author{J. Hugo Dil}
\affiliation{Institute  of  Physics,  \'Ecole  Polytechnique  F\'ed\'erale  de  Lausanne  (EPFL),  CH-1015  Lausanne,  Switzerland}%
\affiliation{Spectroscopy of Quantum Materials Group, Paul Scherrer Institute, Villigen, Switzerland}%
\author{Dharmalingam Prabhakaran}
\affiliation{Department of Physics, University of Oxford, Clarendon Laboratory, Oxford OX1 3PU, United Kingdom}
\author{Henrik M. R{\o}nnow}
\affiliation{Institute of Physics, Ecole Polytechnique Fédérale de Lausanne (EPFL), CH-1015 Lausanne, Switzerland}
\author{Fernando de Juan}
\affiliation{Donostia International Physics Center, P. Manuel de Lardizabal 4, 20018 Donostia-San, Sebastian, Spain}
\affiliation{IKERBASQUE, Basque Foundation for Science, Maria Diaz de Haro 3, 48013 Bilbao, Spain}
\author{Maia G. Vergniory}
\affiliation{Donostia International Physics Center, P. Manuel de Lardizabal 4, 20018 Donostia-San, Sebastian, Spain}
\affiliation{Max Planck Institute for Chemical Physics of Solids, 01187 Dresden, Germany}
\author{Andrew T. Boothroyd}
\affiliation{Department of Physics, University of Oxford, Clarendon Laboratory, Oxford OX1 3PU, United Kingdom}
\date{\today}
\begin{abstract}
In the rapidly expanding field of topological materials there is growing interest in systems whose topological electronic band features can be induced or controlled by magnetism. Magnetic Weyl semimetals, which contain linear band crossings near the Fermi level, are of particular interest owing to their exotic charge and spin transport properties. Up to now, the majority of magnetic Weyl semimetals have been realized in ferro- or ferrimagnetically ordered compounds, but a disadvantage of these materials for practical use is their stray magnetic field which limits the minimum size of devices. Here we show that Weyl nodes can be induced by a helical spin configuration, in which the magnetization is fully compensated. Using a combination of neutron diffraction and resonant elastic x-ray scattering, we find that EuCuAs develops a planar helical structure below $T_\textrm{N}$ = 14.5\,K which induces Weyl nodes along the $\Gamma$--A high symmetry line in the Brillouin zone. 
\end{abstract}
\maketitle
\section{Introduction}
Weyl semimetals (WSMs) are crystalline solids characterized by points in momentum space where singly degenerate bands cross. These points, called Weyl nodes, are extremely robust against perturbations due to their non-trivial topology. In recent years, WSMs have garnered significant attention from both theoretical and experimental perspectives due to their potential to host relativistic charge carriers which mimic the behavior of massless fermions, and the associated exotic transport properties~\cite{Armitage2018,YanFelser2017,CayssolFuchs2021,LvQianDong2021} 

The realization of a WSM requires the breaking of either (or both of) inversion symmetry ($\mathcal{P}$) or time-reversal symmetry ($\mathcal{T}$). The latter approach is of particular interest because it provides a route to greater control over the topology of electronic bands via magnetism or magnetic fields~\cite{Bernevig_2022_review,Tokura2019}. Ferromagnetic WSMs are an obvious choice of system to work with, but their instability at small scales due to stray fields limits their potential for downscaled applications~\cite{Liu_2018_CSS,Destraz_2020_PrAlGe,Kim_2018_FGT,Ilya_2019_CMG}. Antiferromagnetic (AFM) WSMs are more promising in terms of stability, but most AFM structures do not lift the double degeneracy of the bands because although collinear AFM order
 breaks $\mathcal{T}$, it usually preserves the combination $\mathcal{P}\times \mathcal{T}$, which prevents any Dirac points from splitting into pairs of Weyl nodes. 

\begin{figure}[b!]
\includegraphics[width=0.49\textwidth]{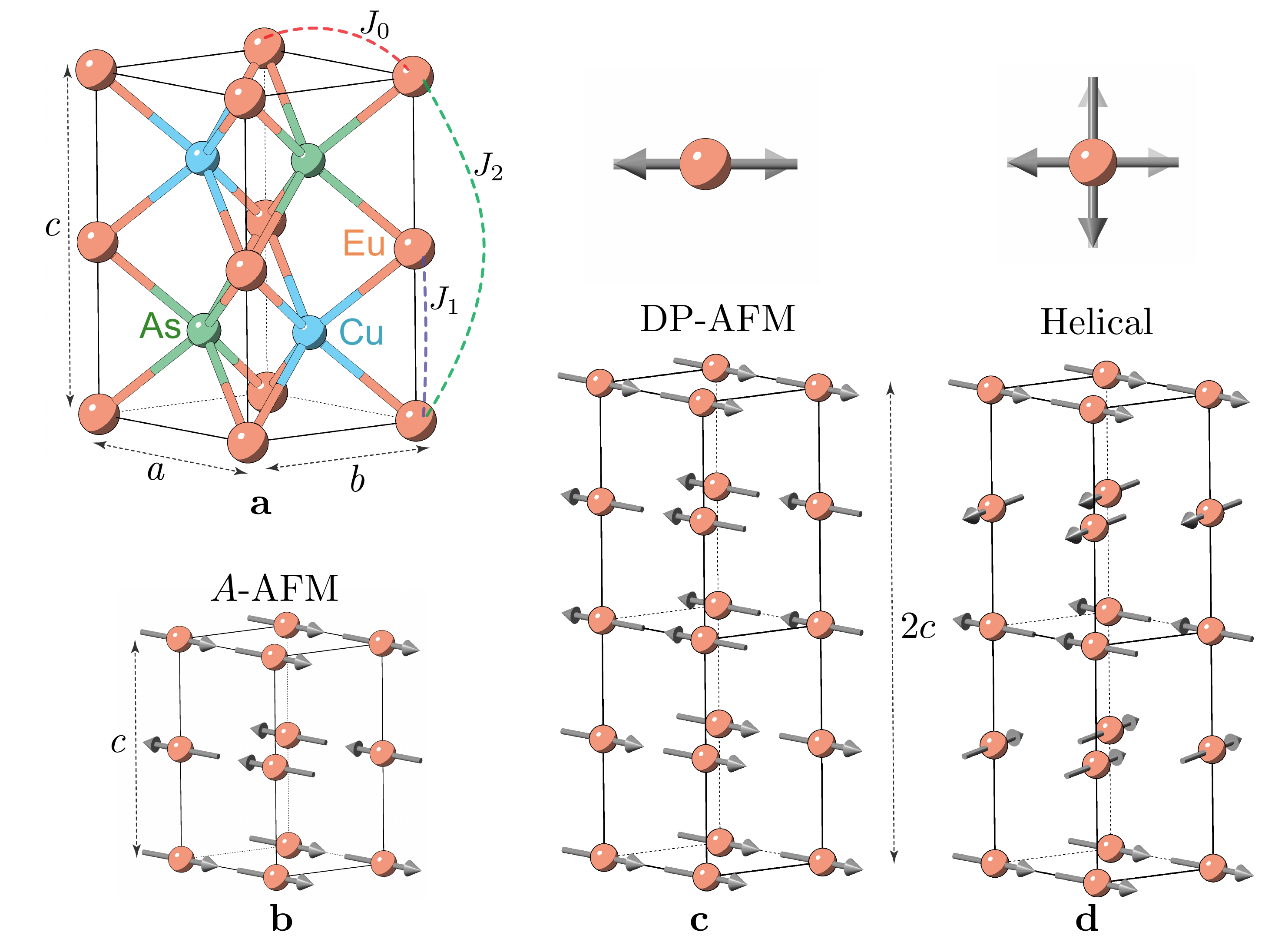}
\caption{\label{fig:Figure_1} \textbf{a} The hexagonal unit cell of \eca{} is described by the centrosymmetric $P6_3/mmc$ space group, with Cu-As layers hosting topological fermions sandwiched between magnetic Eu layers. \textbf{b}--\textbf{d} Possible Eu spin configurations in EuCuAs: \textbf{b} $A$-type AFM order, with a magnetic propagation vector of $\textbf{q}_\textrm{m}=(0,0,0)$; \textbf{c} double-period antiferromagnetic structure (DP-AFM); \textbf{d} helical spin arrangement (right-handed chirality is shown). The latter two structures possess a magnetic propagation vector of $\textbf{q}_\textrm{m} = (0,0,\frac{1}{2})$, i.e. with a doubling of the unit cell along the crystal $c$ axis.}
\end{figure}

An alternative approach is to look for AFM systems with non-collinear spin arrangements that do not have $\mathcal{P}\times \mathcal{T}$ symmetry. An example is the Mn$_3X$ compounds ($X = $ Sn, Ge)  which have a type of chiral $120^\circ$ AFM structure that supports Weyl nodes~\cite{KublerFelser2017}, but whose electronic bands near the Fermi level are strongly broadened and renormalized and therefore difficult to probe experimentally~\cite{Kuroda2017}. On the other hand, the non-collinear  configurations of Eu spins in centrosymmetric materials such as EuCo$_2$P$_2$~\cite{Reehius_1992_ECP}, EuNi$_2$As$_2$~\cite{Jin_2019_ENA}, EuZnGe~\cite{Kurumaji_2022_EZG} and EuCuSb~\cite{Takahasi_2020_ECS}
may provide a way to lift the band degeneracy without introducing significant electronic correlations. Sizable band splittings of order 0.1\,eV are possible due to the large exchange coupling between localized Eu $4f$ states and the conduction electrons~\cite{soh_ideal_2019}. Calculation of the electronic structure of these Eu compounds by density functional theory (DFT) is problematic due to the large magnetic supercells, and up to now calculations have either not been attempted or do not include the full non-collinear incommensurate magnetic structure. Hence, a simpler system with a commensurate non-collinear AFM order is desired for further investigation. 

    In this study we consider \eca{}, which belongs to the centrosymmetric Eu$TX$($T =$ Cu, Ag, Au; $X = $ P, As, Sb, Bi) family of materials~\cite{Mewis1978,Tomuschat1984,du2015dirac,tong2014magnetic,nakamura_thermoelectric_2023,EuAgAs_DFT_Magnetization,Jin_EuAgAs_DFT,EuAuAs_DFT_Magnetization,EuCuP_Expt,EuCuBi_DFT} whose unit cell is shown in Fig.~\ref{fig:Figure_1}\textbf{a}. We establish that below the Néel temperature, $T_\mathrm{N} = 14.5$\,K, the Eu moments are aligned ferromagnetically in the basal plane and exhibit a helical spin arrangement along the $c$ axis, with a magnetic propagation vector $\textbf{q}_\textrm{m}=(0,0,\tau)$, $\tau \simeq 0.5$. The magnetic interactions are found to be highly two-dimensional, and the helix is most likely stabilized by frustration in the out-of-plane couplings. Our \textit{ab initio} DFT calculations demonstrate that the helical spin structure in EuCuAs generates Weyl nodes.

\begin{figure}[t!]
\includegraphics[width=0.49\textwidth]{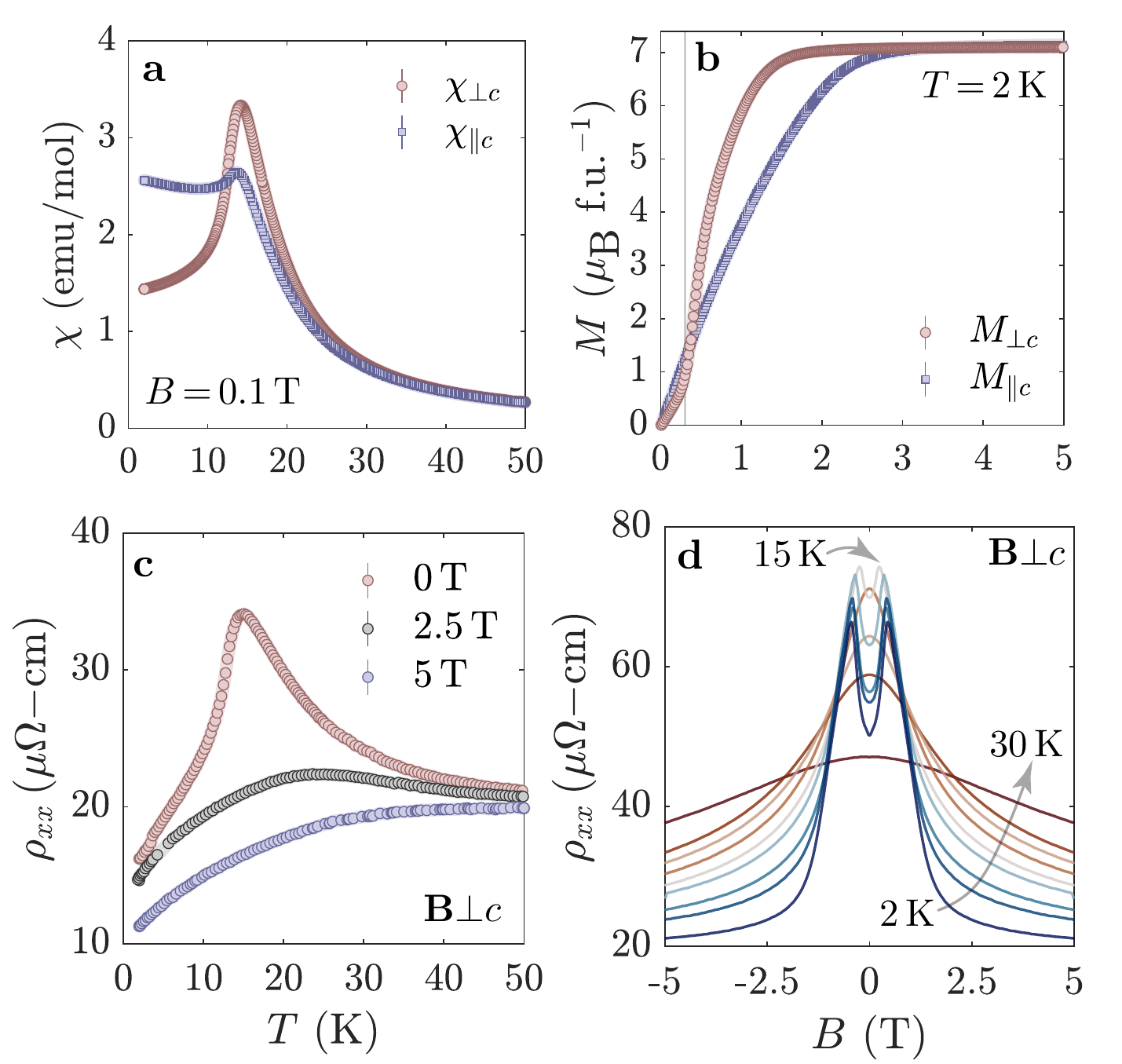}
\caption{\label{fig:Figure_2} \textbf{Magnetic and transport properties of \eca{}.} Figure 1: \textbf{a} Magnetic susceptibility curves for \eca{} in the $\textbf{B}$$\perp$$c$ and $\textbf{B}||c$ configurations show an anomaly at $T_\mathrm{N}$=14.5\,K. \textbf{b} Magnetization curves as a function of applied field along and perpendicular to $c$ at fixed temperature of $T$=2\,K. \textbf{c} Temperature dependence of resistivity ($\rho_{xx}$) at various fixed field strengths ($|\textbf{B}|$\,=\,0, 2.5, 5\,T; $\textbf{B}$$\perp$$c$). \textbf{d} Magnetic field dependence of $\rho_{xx}$ at various fixed temperatures ($T$ = $2-30$\,K). }
	\end{figure}

\section{Results}
\subsection{Bulk properties}

\begin{figure*}[t!]
\includegraphics[width=0.99\textwidth]{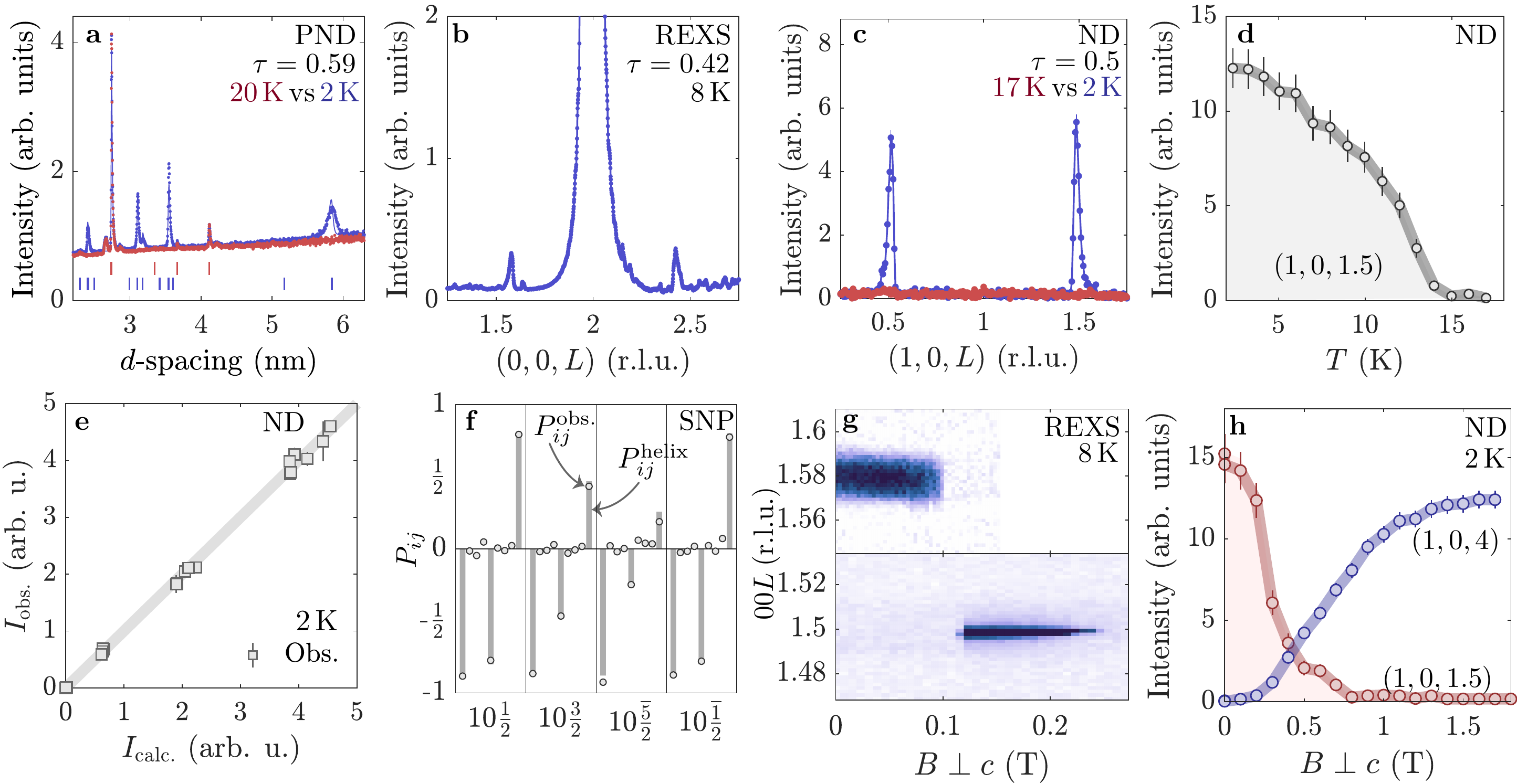}
\caption{\label{fig:Figure_3} \textbf{Unravelling the magnetic order of \eca{} with neutrons and x-rays.} \textbf{a} Powder neutron diffraction (PND). Red and blue tick marks indicate structural and magnetic Bragg peaks. \textbf{b} Resonant elastic X-ray scattering (REXS). The peaks at $L = 1.58$ and $2.42$ are magnetic Bragg peaks. \textbf{c} Single-crystal neutron diffraction (ND). The peaks at $L = 0.5$ and $1.5$ are magnetic Bragg peaks. \textbf{d} Integrated intensity of the \textbf{Q}=$(1,0,1.5)$ magnetic reflection measured with ND. \textbf{e} Refinement of single-crystal ND data, comparing the observed and calculated magnetic Bragg peak intensities for the helical structure. \textbf{f} Comparison of observed ($P_{ij}^\mathrm{obs.}$) and calculated ($P_{ij}^\mathrm{helix}$) full polarization matrices for four magnetic reflections. For each reflection, the nine elements of the polarization matrix $P_{ij}$ from left to right correspond to $ij$ = $xx$, $xy$, $xz$, $yx$, $yy$, $yz$, $zx$, $zy$ and $zz$. The data were recorded at $T = 2$\,K. \textbf{g}, \textbf{h} Field dependence of peaks observed with REXS and ND, respectively.}
\end{figure*}
 
In Figure~\ref{fig:Figure_2}\textbf{a}, we present the temperature-dependent susceptibility of \eca{} measured with applied field parallel and perpendicular to the $c$ axis. The peak observed in both field directions at $T_\mathrm{N}=14.5$\,K indicates the onset of AFM order of the Eu magnetic sublattice. The susceptibility curves and ordering temperature are consistent with earlier measurements reported in~\cite{tong2014magnetic}. Below $T_\mathrm{N}$, it can be seen that the magnetic susceptibility along the $c$ axis ($\chi_{\|c}$) is  greater than that perpendicular to the $c$ axis ($\chi_{\perp c}$) at low temperatures, suggesting that the Eu magnetic moments lie in the $ab$ plane.  

Figure~\ref{fig:Figure_2}\textbf{b} displays field-dependent magnetization curves measured at $T=2$\,K for the two field directions. When the field is applied along the $c$ axis ($\textbf{B}\|c$), the magnetization increases smoothly and reaches saturation above a field of $B_{\|c}^\mathrm{sat.}\simeq 2.5$\,T. This behavior is consistent with an increasing tilt of Eu magnetic moments out of the plane until full polarization is reached along the field direction. The saturation magnetization, $M^\mathrm{sat} = 7.0(1) \mu_\mathrm{B}$\,f.u.$^{-1}$, agrees well with the expected value  $g_J J\mu_\textrm{B}$ for a single divalent Eu$^{2+}$ ion ($4f^7$, $L=0$, $S=7/2$ and $g_J=2$) per formula unit.

In the $\textbf{B}$$\perp$$c$ field configuration we observe a kink in the magnetization at around $B_\mathrm{t}\sim 0.3$\,T, consistent with previous measurements~\cite{tong2014magnetic}, 
which is not present in the $\textbf{\textit{B}}\|c$ data (see Fig.~\ref{fig:Figure_2}\textbf{b}). Beyond this point, the magnetization increases rapidly and saturates at a field of $B_{\perp c}^\mathrm{sat}\simeq$1.5\,T. The anomaly in the magnetization curve at $B_\mathrm{t}$ suggests a metamagnetic transition in which there is a sudden rearrangement of the Eu moments within the $ab$ plane.

To investigate the relationship between charge transport and Eu magnetism in \eca, we plot the resistivity ($\rho_{xx}$) as a function of temperature for three different fields (Fig.~\ref{fig:Figure_2}\textbf{c}). In the absence of a magnetic field, we observe that $\rho_{xx}$ increases on cooling and reaches a maximum at $T_\mathrm{N}$, before dropping sharply as the temperature approaches 2\,K.  We also find that the resistivity peak can be suppressed by an applied magnetic field, as shown in the measurements at $B = 2.5$\,T and 5\,T (Fig.~\ref{fig:Figure_2}\textbf{c}, \textbf{d}). These observations suggest  that the enhanced resistivity at $T_\mathrm{N}$ is due to the scattering of charge carriers from cooperative Eu spin fluctuations,  which start to build up at temperatures of order 50\,K and reach a maximum amplitude at $T_\mathrm{N}$. Spin fluctuations are quenched by a magnetic field comparable to $B^\textrm{sat}$, and this likely accounts for the reduction in the resistivity at $T_\mathrm{N}$ with field observed in Fig.~\ref{fig:Figure_2}\textbf{c}.  

The $\rho_{xx}$ measurements as a function of field applied perpendicular to the  $c$-axis are plotted in Fig.~\ref{fig:Figure_2}\textbf{d}. At $T=2$\,K, the resistivity increases up to $B_\mathrm{t} = 0.3$\,T, above which the resistivity decreases sharply with field. As the temperature is increased, the low-field anomaly in the resistivity becomes washed out and is completely suppressed at temperatures above $T_\mathrm{N}$. Negative magnetoresistance is observed at all temperatures for fields above $B^\textrm{sat}$ (Fig.~\ref{fig:Figure_2}\textbf{d}). 

\subsection{Magnetic diffraction with  neutrons and x-rays}

We investigated the magnetic structure of the Eu moments below $T_\mathrm{N}$ in \eca\ using powder neutron diffraction (PND), single crystal neutron diffraction (ND), and resonant X-ray magnetic scattering (REXS), as shown in Fig.~\ref{fig:Figure_3}. The PND measurement (Fig.~\ref{fig:Figure_3}\textbf{a}) revealed magnetic diffraction peaks below $T_\mathrm{N}$  which could be indexed with a magnetic propagation vector $\textbf{q}_\textrm{m} = (0, 0, \tau)$ with $\tau = 0.591(1)$ reciprocal lattice units (r.l.u.). This means that the Eu spins align ferromagnetically within the hexagonal planes and have an incommensurate period along the $c$ axis. The refinements gave good agreement with the data for the  proper screw (planar helix) structure shown in Fig.~\ref{fig:Figure_1}\textbf{d}, in which the spins lie in the $ab$ plane and rotate around the $c$ axis from layer-to-layer.  The statistics place an upper bound of 3$^\circ$ on any out-of-plane tilt. 

We also found a magnetic propagation vector of the form $\textbf{q}_\textrm{m} = (0, 0, \tau)$ in the REXS study of a EuCuAs single crystal, but this time we observed $\tau = 0.42(1)$\,r.l.u., Fig.~\ref{fig:Figure_3}\textbf{b}. We note that because there are two Eu layers per unit cell, magnetic Bragg peaks are observed at positions $(H,K,2L\pm \tau)$, where $H, K, L$ are integers.

The different values of $\tau$ found in the PND and REXS measurements indicate that the period of the helix can vary from sample to sample, perhaps due to small differences in composition. This finding was reinforced in the single-crystal ND study in which we used a different single crystal sample and found yet another value, $\tau=0.50(1)$. Figure~\ref{fig:Figure_3}\textbf{d} shows the temperature dependence of the integrated intensity of a peak observed at $\textbf{q} = (1,0,1.5)$, which demonstrates an order-parameter-like dependence below $T_\mathrm{N} = 14.5(5)$\,K. This temperature coincides with the peak in the susceptibility (Fig.~\ref{fig:Figure_2}\textbf{a}), and also agrees with  $T_\mathrm{N}$ values found in the PND and REXS measurements, confirming the magnetic origin of the $\tau = 0.5$ family of peaks.

Symmetry analysis for the paramagnetic group $P6_3/mmc$ with propagation vector $\textbf{q}_\textrm{m} = \left(0,0,\frac{1}{2}\right)$ shows that the magnetic group formed by the Eu spin components decomposes into two irreducible representations (irreps), $\Gamma = \Gamma_2 + \Gamma_3$. Of these, $\Gamma_2$ describes a pair of longitudinal spin-density wave structures with the spins pointing along the $c$ axis. This structure can immediately be ruled out  because  magnetic peaks of the form $(0,0,2L\pm \frac{1}{2})$ were observed. The $\Gamma_3$ irrep describes four domains of the planar helix shown in Fig.~\ref{fig:Figure_1}\textbf{d}. We also considered the DP-AFM structure shown in Fig.~\ref{fig:Figure_1}\textbf{c}. This structure is not described by one of the irreps, so based on Landau's theory it is not expected to form in a continuous magnetic phase transition. Nevertheless, we consider it because after averaging over an equal population of domains related by $\pm 120^\circ$ in-plane rotations of the Eu moments about the $c$ axis the magnetic Bragg peak intensities are identical with those of the helical structure. Additionally, the DP-AFM structure with canted spins is found to be stabilized by an in-plane magnetic field (see below), so it is important to rule out this structure in zero field.


 The single-crystal ND data in zero field are described well by both structures assuming equal domain populations, with $\chi_\textrm{r}^2 = 6.51$ ($\chi_\textrm{r}^2$ is the usual goodness-of-fit statistic normalized to the number of degrees of freedom).  
 Figure~\ref{fig:Figure_3}\textbf{e} shows the good agreement between the observed and calculated peak intensities for the helical structure. However, the large shape-dependent corrections needed to account for the severe neutron absorption of Eu in the sample (absorption cross-section 4,530\,b at 1.8\,\AA \ wavelength) make a more detailed structural analysis from this data unreliable.

To provide a more stringent investigation of the zero-field magnetic order in the EuCuAs sample with $\tau = 0.5$ we turned to spherical neutron polarimetry (SNP). In the SNP technique~\cite{Boothroyd_book}, the information on the magnetic structure is obtained from the polarization matrix $P_{ij}$. The components of $P_{ij}$ are determined from intensity ratios (see Methods) and so do not depend on sample attenuation, which is an advantage for samples with strong neutron absorption like EuCuAs. As with unpolarized neutron diffraction, SNP cannot distinguish between the helical and DP-AFM  models when the latter is averaged over equal populations of equivalent domains. However, a sample that is field-cooled through $T_\textrm{N}$ into the DP-AFM phase is expected to preferentially contain domains in which the spins lie perpendicular to the field. With this in mind, we cooled the sample from 25\,K to 2\,K in a field of 1\,T applied parallel to the $b$ axis, before removing the field, inserting the cryostat into the SNP device and measuring the $P_{ij}$ at ten magnetic peak positions in zero field. A fit to a single domain of the DP-AFM structure in which all spins are perpendicular to the field gave a poor agreement with $\chi_\textrm{r}^2 = 649$. Allowing the domain populations to vary, we found that the data constrain the DP-AFM domain imbalance to be less than 5\%. In contrast, a refinement of the helical magnetic structure against the  $P_{ij}$ gave an excellent fit, with $\chi_\textrm{r}^2 = 1.85$, as illustrated in Fig.~\ref{fig:Figure_3}\textbf{f} for four of the reflections.

		
We conclude, therefore, that the magnetic structure of EuCuAs in zero or very small fields is a planar helix with a (sample-dependent) period of approximately four Eu layers, as shown in Fig.~\ref{fig:Figure_1}\textbf{d}. 
Next, we investigate the metamagnetic transition observed when a magnetic field is applied perpendicular to the $c$ axis (Fig.~\ref{fig:Figure_2}\textbf{b}).  Figure~\ref{fig:Figure_3}\textbf{g} shows that with increasing field the incommensurate $(0,0,1.58)$ peak observed at $T = 8$\,K in the REXS experiment changes in position very slightly at first, but then jumps discontinuously at a field of $B \simeq 0.1$\,T  to the commensurate position $(0,0,1.5)$.  The ND crystal has a commensurate propagation vector already at zero field, but with increasing field we observe a switch in the magnetic scattering intensity from $L = $ half-integer to $L = $ integer positions, as illustrated in Fig.~\ref{fig:Figure_3}\textbf{h}. At fields above about 0.2\,T the intensity of the $(1, 0, 1.5)$ reflection  decreases sharply, while the $(1,0,4)$ reflection increases. The crossover occurs at a field of around 0.3\,T at $T = 2$\,K, which corresponds roughly with the metamagnetic transition field $B_\textrm{t}$, see Fig.~\ref{fig:Figure_2}\textbf{b}. The lower value of $B_\textrm{t}$ observed by REXS is most likely explained by the higher temperature of the sample during the measurement.   	
\begin{figure*}[t!]
\includegraphics[width=0.99\textwidth]{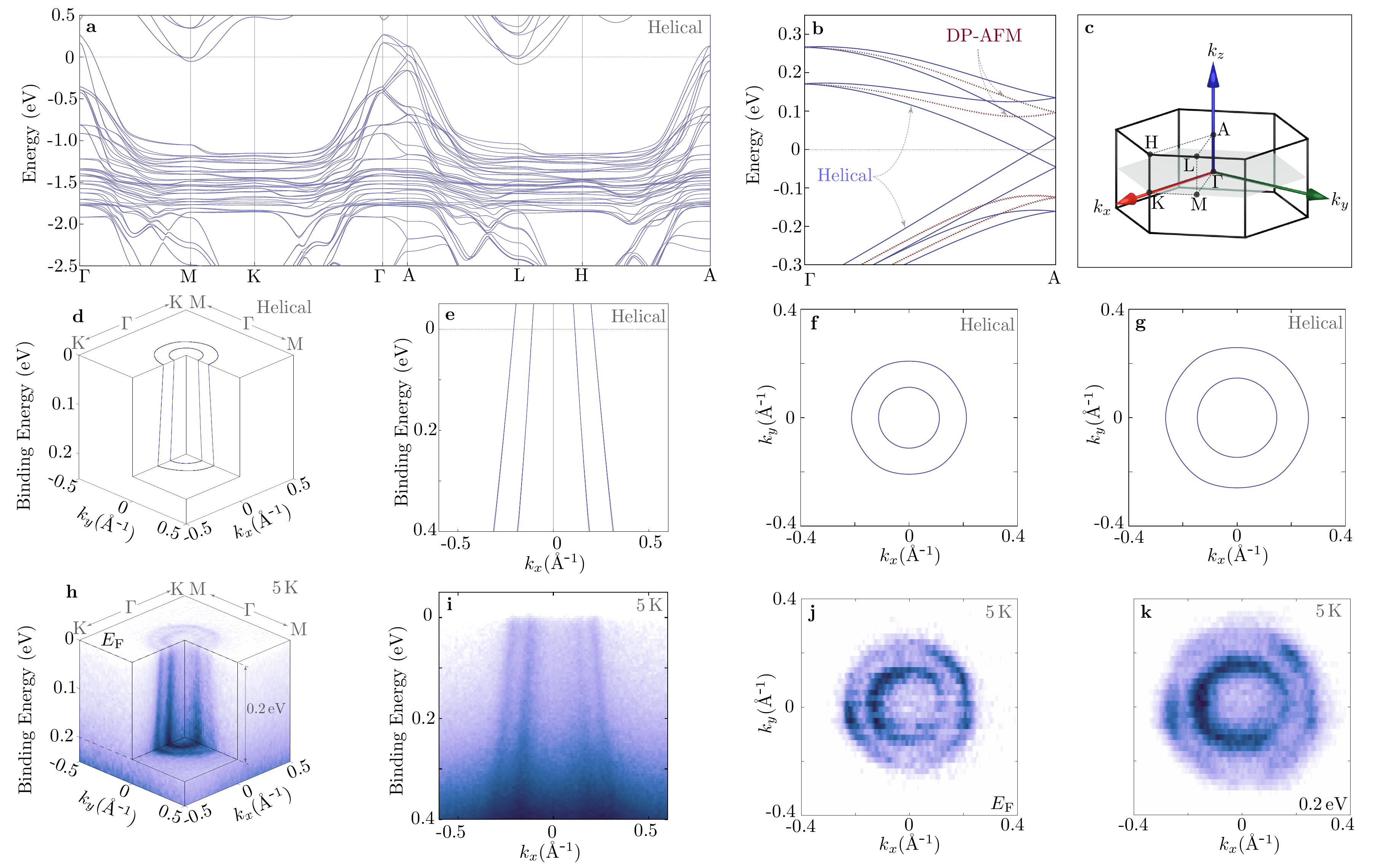}
\caption{\label{fig:Figure_4} \textbf{Comparison between the calculated and measured electronic band structure of \eca{}.} \textbf{a}-\textbf{g} \textit{Ab-initio} electronic structure calculations. \textbf{a} Dispersion along high symmetry lines with helical magnetic order. \textbf{b} Comparison between the dispersion along $\Gamma$-A for a helical magnetic configuration (solid line) and DP-AFM structure (dotted line). \textbf{c} Hexagonal Brillouin zone of \eca{}. \textbf{d}-\textbf{g} Cross-section of the three-dimensional band structure, dispersion along K$-$$\Gamma$$-$K, and constant energy maps at $E_\mathrm{F}$ and 0.2\,eV, respectively. An upward shift of 0.4\,eV has been applied to the calculated bands shown in order to match the ARPES data. The folded bands near $\Gamma$ at about $-0.5$\,eV (unshifted energy) have very little spectral weight and are omitted for clarity (see Supplementary Information). \textbf{h}-\textbf{k} measured ARPES spectrum at $T=5$\,K. }
\end{figure*}

To clarify the arrangement of the Eu moments above the metamagnetic transition field, a total of 316 reflections (both integer and half-integer $L$) were measured at $B=0.4$\,T. The refinement of the model to the observed reflections reveals that at this field the Eu moments form a canted DP-AFM structure, which can be considered as the sum of collinear DP-AFM order and FM order. The DP-AFM component is a single domain with the Eu spins perpendicular to the field. The integer reflections indicate a  FM component along the direction of the applied field ($\textbf{B} \parallel b$). Therefore, considering our results in zero field and in 0.4\,T we conclude that at $B_\mathrm{t} \sim 0.3$\,T the Eu moments reorient within the $ab$ plane from the helical to a canted DP-AFM structure.

At higher fields ($B \geq 0.8$\,T), the integrated intensity of $(1,0,1.5)$ peak remains zero while the $(1,0,4)$ intensity continues to increase, eventually saturating at $B \simeq 1.5$\,T, where the magnetization also plateaus (see Fig.~\ref{fig:Figure_2}\textbf{b}). Thus, the behavior of the magnetic intensities above $B_\mathrm{t}$ is consistent with the gradual canting of the Eu moments along the direction of the applied field into a fully polarized state.

\subsection{Electronic band structure}
To clarify the nature of the charge carriers in \eca\, below $T_\mathrm{N}$, we plot in Fig.~\ref{fig:Figure_4}\textbf{a} the electronic band structure calculated with a commensurate period-4 ($\tau = 0.5$) helical arrangement of the Eu magnetic moments (Fig.~\ref{fig:Figure_1}\textbf{d}). Our calculations indicate that the bands that give rise to charge transport (i.e. close to the Fermi energy, $E_\mathrm{F}$) are dominated by states with Cu and As character, while the Eu $4f$ bands which are responsible for magnetism reside $\sim$1.2\,eV below $E_\mathrm{F}$ (see Supplementary Information). The bands are singly degenerate throughout the hexagonal Brillouin zone (Fig.~\ref{fig:Figure_4}\textbf{c}), with the spin-splitting driven by the chiral spin configuration which breaks all of the mirror symmetries of the $P6_3/mmc$ space group. As such, band crossings can be topologically non-trivial and can host Weyl fermions if the nodes are close to the Fermi energy ($E_\mathrm{F}$). Figure~\ref{fig:Figure_4}\textbf{b} shows the electron energy dispersion along $k_z$ from the $\Gamma$ to A high symmetry points of the hexagonal Brillouin zone. We find two Weyl crossings that are created by the helical magnetic order, located at the A high-symmetry point about 0.18\,eV above and below $E_\mathrm{F}$.
	
As a comparison, we also plot in the same energy--momentum window the calculated electronic band structure with the DP-AFM configuration which we have considered but ruled out in the previous section (Fig.~\ref{fig:Figure_4}\textbf{b}). Unlike with the the helical Eu magnetic order, the electronic bands in the DP-AFM structure do not manifest Weyl points at the A point because the bands that cross are doubly degenerate. 
The Kramers degeneracy is preserved because the DP-AFM structure possesses $m_z \times \mathcal{T}$ symmetry which leaves $k_z$ invariant, where the mirror $m_z$ is the CuAs plane between two neighboring Eu atoms with parallel spins. 
	
In order to validate our \textit{ab initio} calculations of the electronic bands, we compare them with angle-resolved photoemission spectroscopy (ARPES) measurements. We find very good agreement providing the calculated bands are shifted up in energy by approximately 0.4\,eV, indicating that this sample is slightly hole-doped. Figure~\ref{fig:Figure_4}\textbf{d} plots the three-dimensional cross-section of the calculated electronic dispersion, which shows two concentric conical bands centered at the $\Gamma$ point, in excellent agreement with the measured spectrum obtained at $T$=\,5\,K (Fig.~\ref{fig:Figure_4}\textbf{h}). The calculated dispersion along the $K$$-$$\Gamma$$-$$K$ high-symmetry direction reveals two pairs of linearly dispersing bands (Fig.~\ref{fig:Figure_4}\textbf{e}), which we also find in the measured spectrum in Fig.~\ref{fig:Figure_4}\textbf{i}.

Moreover, the calculated Fermi surface (binding energy $E = 0$) displays two concentric circular hole-like pockets centered at the $\Gamma$ point (Fig.~\ref{fig:Figure_4}\textbf{f}). The constant-energy cut at $E =0.2$\,eV, shown in Fig.~\ref{fig:Figure_4}\textbf{g}, also has two concentric circular bands but are larger than that at the Fermi energy. Correspondingly, the measured spectrum at $E_\mathrm{F}$ and at 0.2\,eV (Fig.~\ref{fig:Figure_4}\textbf{j},\textbf{k}) are in good agreement with the dispersion of the calculated bands.
	
\subsection{Spin Hamiltonian}

In this section we shall develop an approximate spin Hamiltonian to describe the magnetic behavior of EuCuAs. Because the spins on Eu are large ($S=\frac{7}{2}$) and localised we expect that a semiclassical mean-field treatment should account for the static magnetic properties, and linear spin-wave theory should give a good description of the magnetic excitations.

Helical magnetic structures in which the spins align ferromagnetically within the layers but rotate from layer-to-layer around the axis of the helix are not uncommon. In materials with inversion symmetry, such that the Dzyaloshinskii--Moriya interaction is absent, they are usually considered to arise from competing exchange interactions. The simplest models are based on the Hamiltonian
\begin{align}
\mathcal{H} = \sum_{\langle ij\rangle} -J_{ij}\textbf{S}_i\cdot \textbf{S}_j + D(S_i^z)^2 + g\mu_\textrm{B}\textbf{S}_i\cdot \textbf{B},
\label{eq:Hamiltonian}
\end{align}
which has been analysed in great detail~\cite{Yoshimori1959,Nagamiya1962,Kitano1964,Nagamiya1967,Robinson1970,Johnston2017}. In the Heisenberg coupling term it will be sufficient to include three exchange constants, $J_0$, which couples neighboring spins in the $ab$ plane, and $J_1$ and $J_2$, which couple nearest and next-nearest neighbor spins along the $c$ axis (Fig.~\ref{fig:Figure_1}\textbf{a}). With $D>0$, the second term ensures easy-plane anisotropy, as observed. Later we discuss the possibility of small additional terms to $\mathcal{H}$.

 With zero applied magnetic field ($\textbf{B} = 0$), it is well known that the ground state magnetic structure obtained from Eq.~(\ref{eq:Hamiltonian}) in the mean-field approximation is a planar helix with turn-angle between spins of 
  \begin{align}
     \theta = \cos^{-1}\left(-\frac{J_1}{4J_2}\right).
    \label{eq:helix-prop-vec} 
    \end{align}
The helix is the stable solution for $J_2 < 0$ and $|J_1| < |4J_2|$. In EuCuAs, which has two Eu layers per unit cell along the $c$ axis, the  propagation vector of the helix is $\textbf{q}_\textrm{m} = (0,0,\tau)$\,r.l.u. with $\tau = \theta/\pi$.

In general, the behavior of a planar helix in a field applied perpendicular to the axis of the helix can be complex, with several different phases predicted as a function of field and helical turn-angle~\cite{Yoshimori1959,Nagamiya1962,Kitano1964,Robinson1970,Johnston2017}.  However, for $\theta \simeq\pi/2$, as established here in EuCuAs, it was found that a transition takes place at an intermediate field $B_\textrm{t}$ to a canted double-period AFM structure~\cite{Nagamiya1962}. In other words, at $B_\textrm{t}$ the propagation vector jumps to $\textbf{q}_\textrm{m} = (0,0,\frac{1}{2})$ corresponding to a turn-angle of $\pi/2$. This is precisely what we have observed in EuCuAs, Fig.~\ref{fig:Figure_3}\textbf{g}. By applying the mean-field approximation to the Hamiltonian (\ref{eq:Hamiltonian}), and neglecting any distortion of the ideal helix for $B_{\perp c}<B_\textrm{t}$, we find that $B_\textrm{t}$ and the saturation fields are given by
\begin{align}
B_\textrm{t} & = |J_1|\sqrt{\frac{J_1+2J_2}{2J_2}}\frac{S}{g \mu_\textrm{B}}
\label{eq:transition-field-ab}\\[5pt]
B_{\perp c}^\textrm{sat} & = -2(J_1+2J_2)\frac{S}{g \mu_\textrm{B}},
\label{eq:sat-field-ab}\\[5pt]
B_{\parallel c}^\textrm{sat} & = 2\left\{D -J_1(1-\cos\pi\tau)-J_2(1-\cos 2\pi\tau)\right\}\frac{S}{g \mu_\textrm{B}},
\label{eq:sat-field-c}
\end{align}
with the requirement that $J_1+2J_2 < 0$ in addition to $J_2<0$.

We must emphasise that Eqs.~(\ref{eq:transition-field-ab})--(\ref{eq:sat-field-ab}) are only approximate due to the neglect of the distortion of the helix at small in-plane fields, as well as the possibility of a fan structure at fields approaching $B_{\perp c}^\textrm{sat}$.  We have also neglected in-plane anisotropy, which even at very low levels can significantly change the phase stability depending on the direction of the field relative to the easy in-plane directions~\cite{Kitano1964,Robinson1970}. Nevertheless, Eqs.~(\ref{eq:transition-field-ab})--(\ref{eq:sat-field-c}) can give us an estimate of the Hamiltonian parameters for the samples with incommensurate propagation vectors. 

We first use Eq.~(\ref{eq:helix-prop-vec}) to determine $J_1/J_2$, and then Eqs.~(\ref{eq:sat-field-ab}) and (\ref{eq:sat-field-c}) together with the observed $B_{\perp c}^\textrm{sat}$ and $B_{\parallel c}^\textrm{sat}$ to determine $J_1$, $J_2$ and $D$.  The results are given in Table~\ref{Table:H-params}. As a consistency check, we also calculate $B_\textrm{t}$ from the obtained $J_1$ and $J_2$ values using Eq.~(\ref{eq:transition-field-ab}). We find $B_\textrm{t} \simeq 0.46$\,T ($\tau=0.42$) and $B_\textrm{t} \simeq0.29$\,T ($\tau=0.59$), in satisfactory agreement with the experimental value $B_\textrm{t} \simeq 0.3$\,T.

\begin{table}[bt]
\caption{\label{table2}\label{Table:H-params} Estimated exchange and anisotropy parameters (in meV) for EuCuAs. The two sets of parameters are for samples with different incommensurate propagation vectors $\tau$. $J_0$ was determined experimentally for the $\tau=0.59$ sample only.}
\begin{ruledtabular}
\begin{tabular}{ccccc}
$\tau$ & $J_0$ & $J_1$ & $J_2$ & $D$\\
0.42 & $-$ & 0.021 & $-0.021$ & 0.009\\
0.59 & $0.060(3)$ & $-0.0077$ & $-0.0069$ & 0.011\\
\end{tabular}
\end{ruledtabular}
\end{table}

Although the Hamiltonian in Eq.~(\ref{eq:Hamiltonian}) is a reasonable approximation, it cannot describe the sample with $\tau = 0.5$ as it stands. To produce $\tau = 0.5$, Eq.~(\ref{eq:helix-prop-vec}) would require $J_1=0$, in which case Eq.~(\ref{eq:transition-field-ab}) gives $B_\textrm{t}=0$. This would imply that there was no metamagnetic transition and that the period-4 helix was not stable in zero field, both contrary to what we have observed. 
Possible ways in which the Hamiltonian could be modified to stabilise a period-4 helix in zero field include: (1) six-fold in-plane anisotropy, as expected from the crystal structure of EuCuAs, which would stabilise a period-4 helix with the Eu spins aligned along the easy directions in the plane (i.e.~the turn-angle would alternate between $\pi/3$ and $2\pi/3$); (2) a biquadratic interaction $J_\textrm{b}(\textbf{S}_i\cdot \textbf{S}_{i+1})^2$ with $J_\textrm{b} > 0$, which favours a turn-angle of $\pi/2$ (Ref.~\onlinecite{Koshelev2022}); (3) metallic (RKKY) exchange. 

\begin{figure}[t!]
\includegraphics[width=0.49\textwidth]{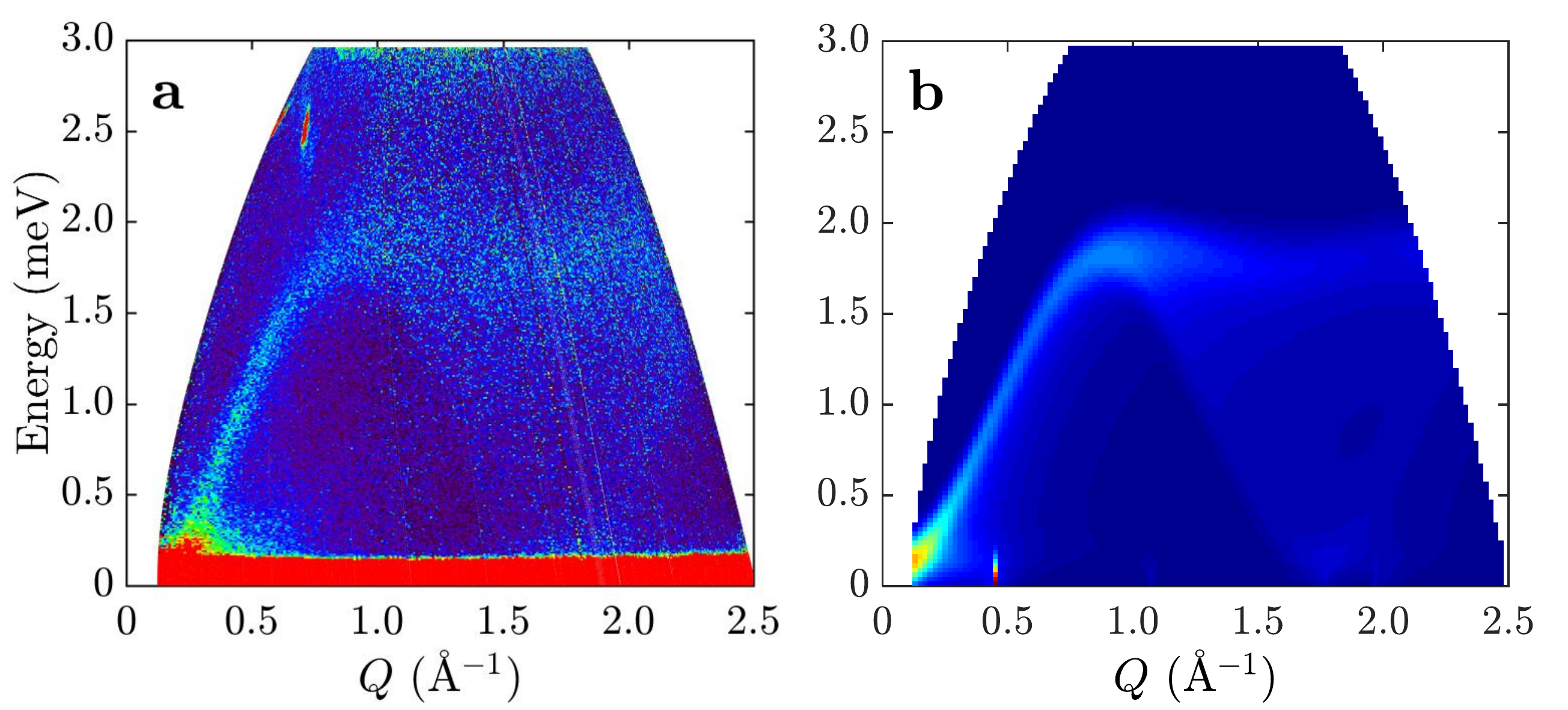}
\caption{\label{fig:Figure_5} \textbf{Spin-wave spectrum of EuCuAs.} \textbf{a} Inelastic neutron scattering data recorded at $T = 2$\,K. \textbf{b} Simulation from linear spin-wave theory with the parameters given in Table~\ref{Table:H-params} for the sample with $\tau = 0.59$.}
\end{figure}

Next, we determine the in-plane nearest-neighbour exchange parameter $J_0$ from the spin-wave spectrum. The inelastic neutron scattering spectrum is reported in Fig.~\ref{fig:Figure_5}\textbf{a}, measured on the same sample as used for the neutron powder diffraction study, Fig.~\ref{fig:Figure_3}\textbf{a}. Despite the punitive neutron absorption of Eu, the spin-wave scattering signal is clearly observable in the neutron scattering data, especially for momentum transfers $Q$ below about 1\,\AA$^{-1}$ where a sharp dispersive mode is observed. 

We simulated the neutron scattering intensity by powder-averaging the theoretical expressions obtained from Eq.~(\ref{eq:Hamiltonian}) by linear spin-wave theory (see Methods). With the Hamiltonian parameters $J_1$, $J_2$ and $D$ obtained above (Table~\ref{Table:H-params}) we find that the inter-layer magnon dispersion has a band width of only 0.1\,meV, and so cannot be resolved from the strong elastic scattering that extends in energy up to about 0.2\,meV (see Fig.~\ref{fig:Figure_5}\textbf{a}). Therefore, the measured spectrum consists almost entirely of the in-plane magnon dispersion.  

The simulation displayed in Fig.~5\textbf{b}, performed with nearest-neighbour exchange parameter $J_0 = 0.060$\,meV and other parameters as given in Table~\ref{Table:H-params}, is seen to agree well with the observed spectrum. Inclusion of in-plane exchange couplings to spins beyond the nearest neighbors did not improve the agreement. Our analysis indicates that the magnitude of the next-nearest-neighbor exchange is less than 10\% of $J_0$. 

The modelling presented in this section has found that the nearest-neighbor intra-layer exchange coupling $J_0$ is between 3 and 8 times larger than the nearest- and next-nearest-neighbor inter-layer couplings $J_1$ and $J_2$ (see Table~\ref{Table:H-params}).  Moreover, each Eu spin is coupled to six other spins by $J_0$, and to only two other spins by $J_1$ and $J_2$. The magnetism in EuCuAs, therefore, is highly two-dimensional. Given this, it is perhaps surprising that the magnetic order is so well correlated in the out-of-plane direction, as reflected in the narrow widths of the peaks in $L$ scans --- see Figs.~\ref{fig:Figure_3}\textbf{b} and \textbf{c}. We estimate the correlation length along the $c$ axis to be approximately 100\,\AA. This suggests that the material is relatively free from chemical disorder of the kind that might disrupt the propagation of the magnetic structure.

\section{Discussion}

Our measurements and analysis provide strong evidence that the Eu spins in EuCuAs adopt a helical spin structure in zero field  below $T_\textrm{N} = 14.5$\,K. The magnetic propagation vector $\textbf{q}_\textrm{m} \simeq (0, 0, 0.5)$  is very different from the propagation vector $\textbf{q}_\textrm{m} = (0, 0, 0)$ of the A-type AFM structure  (Fig.~\ref{fig:Figure_1}\textbf{b}) which had previously been assumed~\cite{tong2014magnetic} and which is often observed in hexagonal layered Eu compounds, e.g.~Refs.~\cite{Rahn_2018_ECA,Soh_2019_ECS,Blawat_2022_EZA,Gui_2019_ESP,Marshall_2021_EMB,Pakhira_2022_EMS}. 
At the same time, helical or helical-like magnetic phases have been observed in a number of other layered Eu compounds, e.g.~Refs.~\cite{Riberolles_2021_EIA,Soh_2023_EIA,Reehius_1992_ECP,Jin_2019_ENA,Sangeetha_2019_ENA,Iida_2019_ERFA,Kurumaji_2022_EZG,Takahasi_2020_ECS}.
As it is difficult to distinguish helical and AFM order from bulk magnetization measurements or \textit{ab initio} calculations it is important to establish the magnetic structure more directly from a microscopic probe of the spin configuration, such as neutron or resonant magnetic x-ray diffraction.

We have found that the magnetic order in EuCuAs has a profound effect on the electronic band topology. The $P6_3/mmc$ space group of EuCuAs has inversion symmetry, so in the paramagnetic phase the bands are doubly degenerate. In the helical phase, however,  inversion and time-reversal symmetries are lost, and as a result the bands become singly degenerate and the band crossings along $\Gamma$--A can be Weyl nodes. 

In EuCuAs, therefore, we have an unusual state of affairs in which a topological phase transition is driven by a magnetic transition from a paramagnet to a helical magnetic order. This differs from most other known magnetic Weyl semimetals, which are induced by ferromagnetism or by externally applied magnetic fields~\cite{Bernevig_2022_review}. Recently, weak helimagnetism was reported in NdAlSi, which is also a magnetic Weyl semimetal~\cite{Gaudet_2021_NdAlSi}. The dominant interactions in NdAlSi favour ferrimagnetic order, but it is proposed that Weyl fermions modify the coupling between Nd ions and generate Dzyaloshinkii--Moriya and Kitaev-type interactions which are responsible for the weak chiral tilting of the moments away from the easy direction.  Hence, the interplay between helical magnetism and electronic topology is very different in the two materials.  In EuCuAs, helical magnetism generates the Weyl state, whereas in NdAlSi, the Weyl state generate helical magnetism.

The main outcome of this work is the discovery that helical magnetic order induces Weyl nodes in EuCuAs.  This finding expands the range of systems in which magnetic order influences the electronic band topology. Of course, EuCuAs is far from ideal because the response of the Weyl fermions is complicated by the presence of several topologically trivial bands at $E_\textrm{F}$. It would be interesting, therefore, to find a simpler system in which study the interplay between Weyl fermions and a subsystem of helically ordered local moments.

\section{Methods}
\textbf{Crystal growth.} Single crystalline \eca\, was grown by the self-flux method, as described in Ref.~\cite{tong2014magnetic}. The quality and structure of the single crystals was checked with laboratory x--rays on a 6--circle diffractometer (Oxford Diffraction) and Laue diffractometer (Photonic Science). Laboratory single crystal x-ray diffraction confirmed the hexagonal Ni$_2$In-type structure $(P6_3/mmc)$. 

\textbf{Magnetization.} Magnetization measurements were performed on a Physical Properties Measurements System (PPMS, Quantum Design) with the vibrating sample magnetometer (VSM) option. The temperature-dependent magnetometry measurements were performed in the temperature range $2 \leq T \leq 50$\,K in an external field of $B = 0.1$\,T in two different field configurations, with $\textbf{B}$ parallel and perpendicular to the crystal $c$ axes, respectively. The field-dependent measurements were performed at $T$=$2$\,K in fields up to 5\,T applied in the same two field directions, that is $\textbf{B}$$\parallel$$c$ and $\textbf{B}$$\perp$$c$. 

\textbf{Magnetotransport.} Magnetotransport measurements were performed on the PPMS with the resistivity option, to shed light on the coupling between charge transport and the different spin-configurations. The field was applied perpendicular to the crystal $c$ axis, in field strengths up to $\mu_0 H = 5$\,T and temperatures down to $T = 2$\,K. 

\textbf{Powder neutron diffraction.}
The polycrystalline sample used for neutron powder diffraction and inelastic neutron scattering was prepared from single crystals which were crushed and finely ground.  Measurements were made on the WISH diffractometer at the ISIS Neutron and Muon Facility. A mass of 1.3\,g of powder was contained in a thin-walled cylindrical can of diameter 3\,mm, made from vanadium. The can was mounted inside a helium cryostat. Data were recorded at temperatures between 20\,K and 2\,K. Magnetic Bragg peaks were observed at temperatures below $T_\textrm{N} = 14.5\pm0.5$\,K (see Supplementary Information).  Rietveld profile refinement was performed with the FULLPROF software suite~\cite{fullprof}.  Data from detector banks 2 to 4 were used for the refinements.  The refined lattice parameters at 20\,K in the space group $P6_3/mmc$ were found to be $a = b = 4.2331(1)$\,{\AA} and $c = 8.2354(2)$\,{\AA}. The Lobanov -- alte da Veiga function was used to model the absorption for the cylindrical sample, but due to the severe neutron absorption of Eu (absorption cross-section $\sigma_\mathrm{abs} = 4,530$\,b at wavelength $\lambda = 1.8$\,{\AA}) the absorption correction is not sufficiently accurate to give reliable thermal parameters, site occupancies or magnetic moment sizes (see Supplementary Information).  
 
\textbf{Single-crystal neutron diffraction.} Single crystal neutron diffraction was performed on the D9 diffractometer at the Institut Laue--Langevin (ILL). Hot neutrons ($\lambda = 0.84$\,{\AA}) were used to reduce the absorption due to Eu. To refine the structure in zero field we collected at total of 229 reflections at a temperature of 2\,K with the sample mounted in a four-circle cryostat. The full dataset comprised 109 structural reflections consistent with the $P6_3/mmc$ space group, and 120 corresponding to the magnetic propagation vector $\textbf{q}_\textrm{m} = (0,0,0.5)$. After averaging symmetry-equivalent reflections, there were a total of 38 unique structural reflections and 40 unique magnetic reflections. Magnetic structure models were refined using the {\it Mag2Pol} software~\cite{QureshiMag2Pol}. Attenuation corrections to the peak intensities (due to the strong neutron absorption of Eu) were calculated according to the length of the neutron path through the crystal, whose shape and dimensions were measured for this purpose. To study how the magnetic structure of \eca{} evolves with an applied magnetic field, we performed single crystal neutron diffraction at various field strengths up to $B = 2.5$\,T in a vertical field cryomagnet. The crystal was mounted with the $b$ axis vertical (parallel to the field direction) so that $h0l$ reflections were accessible in the horizontal scattering geometry.

\textbf{Polarized neutron diffraction.}
Polarized neutrons of wavelength 0.83\,\AA\ were employed on the D3 diffractometer (ILL) in the spherical neutron polarimetry (SNP) set-up with the Cryopad device~\cite{Lelievre2005}. The incident beam was polarized by Bragg reflection from  the (111) planes of a crystal of ferromagnetic Heusler alloy (Cu$_2$MnAl). An erbium filter was placed in the incident beam to suppress half-wavelength contamination. Nutator and precession fields were used to control the direction of the beam polarization.  A $^{3}$He spin filter was used to analyse the polarization of the scattered beam.  Standard corrections for time decay of the filter efficiency were applied based on measurements of the (102) structural Bragg peak. The polarization matrix elements $P_{ij}$ are defined as
\begin{align}
    P_{ij} = \frac{N_{ij} - N_{i\bar{j}}}{N_{ij} + N_{i\bar{j}}},
    \nonumber
\end{align}
where $N_{ij}$ and $N_{i\bar{j}}$ are the number of counts (at a Bragg reflection) when the incident neutron polarization is along $i$ and the scattered polarization is measured parallel and antiparallel to $j$, respectively, with $i,j$ being the principal directions $x, y, z$. The direction $x$ is defined as the direction along the scattering vector $\textbf{\textit{Q}}$, $z$ is perpendicular to the scattering plane, and $y$ completes the right-handed set. 

\textbf{Resonant elastic x-ray diffraction.}
The REXS measurements were performed at the I16 beamline, Diamond Light Source. The incident x-ray photon energy was tuned to the Eu $L_3$ edge so as to benefit from the resonant enhancement of the scattered x-ray intensity from the Eu$^{2+}$ ions. For the zero-field measurements, the diffractometer was set to the vertical geometry and the $\sigma$$\to$$\pi^\prime$ scattering channel, to be mainly sensitive to magnetism which can rotate the linear polarization of the incident x-rays. For the magnetic field dependent measurements, the diffractometer was set to the horizontal scattering configuration to accommodate the vertical field magnet. The $\pi$$\to$$\sigma^\prime$ scattering channel was adopted to suppress the charge scattering and enhance the magnetic scattering.
 
\textbf{Density functional theory.} To clarify the topological nature of the electronic band structure in \eca\, and how it evolves in a magnetic field, we performed density functional theory (DFT) calculations of the electronic band structure using VASP~\cite{VASP1,VASP2} \textit{v.}6.2.1. Projector-augmented wave pseudopotentials in the generalized gradient approximation with Perdew Burke Ernzerhof parametrization (PBE)~\cite{Perdew96} were used. Relativistic pseudo-potentials were used in the calculations to account for the large spin-orbit coupling (SOC) arising from the heavy As ions which might lead to band inversion, with a kinetic energy cutoff of $480$~eV. Furthermore, a Hubbard $U = 5.0$ eV was used to model the strong electron-electron correlations and reproduce the observed binding energy of the localized Eu $4f$ bands (see Supplementary Information). A Monkhorst–Pack \textbf{k}-point sampling mesh of $9\times9\times7$ was used~\cite{PhysRevB.13.5188}.


\textbf{Angle-resolved photoemission spectroscopy.} ARPES of \eca{} was performed on the SIS-ULTRA beamline at the Swiss Light Source (SLS). To determine the $\Gamma$ and A high symmetry point, we performed a $k_z$ dependent scan by varying the incident photon energy from 50\,eV to 150\,eV. Samples were cleaved at $T\sim$15\,K under high vacuum ($\sim10^{-8}$ Torr) and measured at various temperatures between 5\,K and 22\,K at incident photon energy of 74\,eV.  Vertical, horizontal, left-handed and right-handed circular incident photon polarization was used. Spectra obtained at different temperatures and in each polarization channel are reported in the Supplementary Information.

\textbf{Inelastic neutron scattering.} Measurements were performed at the ISIS Neutron and Muon Facility, U.K., on the same 1.3\,g polycrystalline sample as used for neutron powder diffraction.  Initial measurements were performed on the Merlin chopper spectrometer with incident neutron energies between 8 and 100\,meV. The spectra established that the magnetic excitation spectrum did not extend beyond 2\,meV in energy. Subsequent measurements of the same sample were made on the cold-neutron chopper spectrometer LET with neutron incident energies of between 2 and 9\,meV. The data presented in Fig.~\ref{fig:Figure_5} were recorded on LET with 3.7\,meV incident neutrons. During the run, the sample was maintained at a temperature of 2\,K in a pumped helium cryostat.

\textbf{Spin-wave spectrum}. The magnetic excitations of the Hamiltonian Eq.~(\ref{eq:Hamiltonian}) were calculated by linear spin-wave theory (LSWT). Following Yosida and Miwa~\cite{Yosida-Miwa-JAP}, we define local coordinates which rotate with the spins in the helix. Introduction of Holstein--Primakoff operators in the local coordinates followed by Fourier transformation brings the Hamiltonian into the form 
\begin{align}
    \mathcal{H} = \mathcal{H}_0 + \frac{1}{2}\sum_\textbf{q} \textsf{X}^\dag_\textbf{q} \textsf{H}_\textbf{q}\textsf{X}_\textbf{q},
\end{align}
where
\begin{align}
\mathcal{H}_0 & = -NS^2J_{\textbf{q}_\textrm{m}} \\[5pt]
\textsf{X}^\dag_\textbf{q} & = (a^\dag_\textbf{q}, a_{-\textbf{q}}) \\[5pt]
\textsf{H}_\textbf{q} & = \left(\begin{array}{cc}A_\textbf{q} & B_\textbf{q} \\ B_\textbf{q} & A_\textbf{q}\end{array}\right).
\end{align}
$S$ is the spin quantum number, $N$ is the number of lattice points, $a^\dag_\textbf{q}$ is a Fourier-transformed boson creation operator, and
\begin{align}
J_\textbf{q} & = \sum_{\mbox{\small\boldmath $\delta$}\ne 0}J_{\mbox{\boldmath
$\delta$}}\exp({\rm i}{\bf q}\cdot{\mbox{\boldmath$\delta$}})\label{eq:J(q)}
\\[5pt]
A_\textbf{q} & = S\left\{J_{\textbf{q}_\textrm{m}} - \mbox{$\frac{1}{2}$}J_\textbf{q} - \mbox{$\frac{1}{4}$}(J_{{\textbf{q}_\textrm{m}}+\textbf{q}} + J_{{\textbf{q}_\textrm{m}}-\textbf{q}}) + D\right\} \\[5pt]
B_\textbf{q} & = S\left\{ \mbox{$\frac{1}{2}$}J_\textbf{q} - \mbox{$\frac{1}{4}$}(J_{{\textbf{q}_\textrm{m}}+\textbf{q}} + J_{{\textbf{q}_\textrm{m}}-\textbf{q}}) - D\right\}.
\end{align}
The summation in eq.~(\ref{eq:J(q)}) extends over the vectors \boldmath$\delta$ which join a spin to the neighboring spins. Diagonalization of the matrix $\textsf{g}\textsf{H}_\textbf{q}$, where $\textsf{g}$ is the metric tensor
\begin{align}
\textsf{g} & = \left(\begin{array}{cc}1 & 0 \\ 0 & -1\end{array}\right),
\end{align}
gives the spin-wave energy
\begin{align}
\hbar\omega_\textbf{q} = \sqrt{A^2_\textbf{q} - B^2_\textbf{q}}.
\label{eq:SW-dispersion}
\end{align}
Equation~(\ref{eq:SW-dispersion}) was first obtained for a planar helix by Yosida and Miwa~\cite{Yosida-Miwa-JAP}.
The eigenvectors of $\textsf{g}\textsf{H}_\textbf{q}$ are related to the Bose operators via the Bogoliubov transformation
\begin{align}
\left(\begin{array}{c}a_\textbf{q} \\ a^\dag_{-\textbf{q}}\end{array}\right) = \left(\begin{array}{cc}u_\textbf{q} & -v_\textbf{q} \\ -v_\textbf{q} & u_\textbf{q}\end{array} \right) \left(\begin{array}{c}\alpha_\textbf{q} \\ \alpha^\dag_{-\textbf{q}}\end{array}\right),
\label{eq:Bogoliubov}
\end{align}
where
\begin{align}
u_\textbf{q}  = \sqrt{\frac{A_\textbf{q}+\hbar\omega_\textbf{q}}{2\hbar\omega_\textbf{q}}},\ \ \ \ \ \ \ \  
v_\textbf{q} = \sqrt{\frac{A_\textbf{q}-\hbar\omega_\textbf{q}}{2\hbar\omega_\textbf{q}}},
\end{align}
and $\alpha^\dag_\textbf{q}$ creates a magnon with wavevector \textbf{q}.

In the dipole approximation, the inelastic neutron scattering intensity is proportional to~\cite{Boothroyd_book}
\begin{align}
S(\textbf{q},\omega) = f^2(q)\textrm{e}^{-2W} \sum_{\alpha\beta}(\delta_{\alpha\beta}-\hat{q}_{\alpha}\hat{q}_{\beta})S_{\alpha\beta}(\textbf{q},\omega),\label{eq:S(q,w)}
\end{align}
$\alpha, \beta = x, y, z$, where $f(q)$ is the magnetic form factor and $\textrm{e}^{-2W}$ is the Debye--Waller factor. The partial response functions $S_{\alpha\beta}(\textbf{q},\omega)$ contain  the Fourier-transformed spin operators $S_\alpha(\textbf{q})$, $S_\beta(\textbf{q})$. To proceed, these operators are expressed in terms of spin operators in local coordinates and hence in terms of  magnon creation and annihilation operators via eq.~(\ref{eq:Bogoliubov}). The final expressions for magnon creation are
\begin{align}
S_{xx}(\textbf{q},\omega) & = S_{yy}(\textbf{q},\omega) \nonumber\\ & = g^2\mu_\textrm{B}^2\frac{NS}{8}\frac{A_{\textbf{q}\pm \textbf{q}_\textrm{m}} - B_{\textbf{q}\pm \textbf{q}_\textrm{m}}}{\hbar\omega_{\textbf{q}\pm \textbf{q}_\textrm{m}}}\nonumber \\ & \ \ \ \ \times n(\omega)\delta(\omega - \omega_{\textbf{q}\pm \textbf{q}_\textrm{m}})\\[10pt]
S_{zz}(\textbf{q},\omega) & = g^2\mu_\textrm{B}^2\frac{NS}{2}\frac{A_\textbf{q} + B_\textbf{q}}{\hbar\omega_\textbf{q}} \nonumber \\ & \ \ \ \ \times n(\omega)\delta(\omega - \omega_\textbf{q}).
\end{align}
The $\pm$ terms in $S_{xx}/S_{yy}$ are to be summed. $n(\omega) = \{\exp(\hbar\omega/k_\textrm{B}T) - 1)\}^{-1}$ is the boson population factor. The partial response functions with $\alpha \ne \beta$ do not contribute to the unpolarized neutron scattering intensity.

\section{Data availability}
The data presented in this paper is available from the corresponding author upon reasonable request.

\bibliographystyle{unsrt}
\bibliography{ref12.bib}

\begin{thebibliography}{10}

\bibitem{Armitage2018}
N.~P. Armitage, Mele~E. J., and A.~Vishwanath.
\newblock Weyl and {Dirac} semimetals in three-dimensional solids.
\newblock {\em Rev. Mod. Phys.}, 90:015001, 2018.

\bibitem{YanFelser2017}
B.~Yan and C.~Felser.
\newblock Topological materials: {Weyl} semimetals.
\newblock {\em Annu. Rev. Condens. Matter Phys.}, 8:337--354, 2017.

\bibitem{CayssolFuchs2021}
J.~Cayssol and J.~N. Fuchs.
\newblock Topological and geometrical aspects of band theory.
\newblock {\em J. Phys. Mater.}, 4:034007, 2021.

\bibitem{LvQianDong2021}
B.~Q. Lv, T.~Qian, and H.~Ding.
\newblock Experimental perspective on three-dimensional topological semimetals.
\newblock {\em Rev. Mod. Phys.}, 93:025002, 2021.

\bibitem{Bernevig_2022_review}
B.~Andrei Bernevig, Claudia Felser, and Haim Beidenkopf.
\newblock Progress and prospects in magnetic topological materials.
\newblock {\em Nature}, 603(7899):41--51, March 2022.

\bibitem{Tokura2019}
Y.~Tokura, K.~Yasuda, and A.~Tsukazaki.
\newblock Magnetic topological insulators.
\newblock {\em Nat. Rev. Mater.}, 1:126--143, 2019.

\bibitem{Liu_2018_CSS}
Enke Liu, Yan Sun, Nitesh Kumar, Lukas Muechler, Aili Sun, Lin Jiao, Shuo-Ying
  Yang, Defa Liu, Aiji Liang, Qiunan Xu, Johannes Kroder, Vicky Süß, Horst
  Borrmann, Chandra Shekhar, Zhaosheng Wang, Chuanying Xi, Wenhong Wang, Walter
  Schnelle, Steffen Wirth, Yulin Chen, Sebastian T.~B. Goennenwein, and Claudia
  Felser.
\newblock Giant anomalous {Hall} effect in a ferromagnetic kagome-lattice
  semimetal.
\newblock {\em Nat. Phys.}, 14(11):1125--1131, November 2018.

\bibitem{Destraz_2020_PrAlGe}
Daniel Destraz, Lakshmi Das, Stepan~S. Tsirkin, Yang Xu, Titus Neupert,
  J.~Chang, A.~Schilling, Adolfo~G. Grushin, Joachim Kohlbrecher, Lukas Keller,
  Pascal Puphal, Ekaterina Pomjakushina, and Jonathan~S. White.
\newblock Magnetism and anomalous transport in the {Weyl} semimetal {PrAlGe}:
  possible route to axial gauge fields.
\newblock {\em npj Quantum Materials}, 5(1):5, January 2020.

\bibitem{Kim_2018_FGT}
Kyoo Kim, Junho Seo, Eunwoo Lee, K.-T. Ko, B.~S. Kim, Bo~Gyu Jang, Jong~Mok Ok,
  Jinwon Lee, Youn~Jung Jo, Woun Kang, Ji~Hoon Shim, C.~Kim, Han~Woong Yeom,
  Byung Il~Min, Bohm-Jung Yang, and Jun~Sung Kim.
\newblock Large anomalous {Hall} current induced by topological nodal lines in
  a ferromagnetic van der {Waals} semimetal.
\newblock {\em Nat. Mater.}, 17(9):794--799, September 2018.

\bibitem{Ilya_2019_CMG}
Ilya Belopolski, Kaustuv Manna, Daniel~S. Sanchez, Guoqing Chang, Benedikt
  Ernst, Jiaxin Yin, Songtian~S. Zhang, Tyler Cochran, Nana Shumiya, Hao Zheng,
  Bahadur Singh, Guang Bian, Daniel Multer, Maksim Litskevich, Xiaoting Zhou,
  Shin-Ming Huang, Baokai Wang, Tay-Rong Chang, Su-Yang Xu, Arun Bansil,
  Claudia Felser, Hsin Lin, and M.~Zahid Hasan.
\newblock Discovery of topological weyl fermion lines and drumhead surface
  states in a room temperature magnet.
\newblock {\em Science}, 365(6459):1278--1281, 2019.

\bibitem{KublerFelser2017}
J.~K\"{u}bler and C.~Felser.
\newblock Weyl fermions in antiferromagnetic {Mn}$_3${Sn} and {Mn}$_3${Ge}.
\newblock {\em EPL}, 120:47002, 2017.

\bibitem{Kuroda2017}
K.~Kuroda, T.~Tomita, M.-T. Suzuki, C.~Bareille, A.~A. Nugroho, P.~Goswami,
  M.~Ochi, M.~Ikhlas, M.~Nakayama, S.~Akebi, R.~Noguchi, R.~Ishii, N.~Inami,
  K.~Ono, H.~Kumigashira, A.~Varykhalov, T.~Muro, T.~Koretsune, R.~Arita,
  S.~Shin, Takeshi Kondo, and S.~Nakatsuji.
\newblock Evidence for magnetic weyl fermions in a correlated metal.
\newblock {\em Nat. Mater.}, 16:1090, 2017.

\bibitem{Reehius_1992_ECP}
M.~Reehuis, W.~Jeitschko, M.H. Möller, and P.J. Brown.
\newblock A neutron diffraction study of the magnetic structure of
  {EuCo$_2$P$_2$}.
\newblock {\em J. Phys. Chem. Solids}, 53(5):687--690, 1992.

\bibitem{Jin_2019_ENA}
W.~T. Jin, N.~Qureshi, Z.~Bukowski, Y.~Xiao, S.~Nandi, M.~Babij, Z.~Fu, Y.~Su,
  and Th. Br\"uckel.
\newblock Spiral magnetic ordering of the {Eu} moments in {EuNi$_2$As$_2$}.
\newblock {\em Phys. Rev. B}, 99:014425, Jan 2019.

\bibitem{Kurumaji_2022_EZG}
Takashi Kurumaji, Masaki Gen, Shunsuke Kitou, Hajime Sagayama, Akihiko Ikeda,
  and Taka-hisa Arima.
\newblock Anisotropic magnetotransport properties coupled with spiral spin
  modulation in a magnetic semimetal {EuZnGe}.
\newblock {\em Phys. Rev. Mater.}, 6:094410, Sep 2022.

\bibitem{Takahasi_2020_ECS}
Hidefumi Takahashi, Kai Aono, Yusuke Nambu, Ryoji Kiyanagi, Takuya Nomoto,
  Masato Sakano, Kyoko Ishizaka, Ryotaro Arita, and Shintaro Ishiwata.
\newblock Competing spin modulations in the magnetically frustrated semimetal
  {EuCuSb}.
\newblock {\em Phys. Rev. B}, 102:174425, Nov 2020.

\bibitem{soh_ideal_2019}
J.-R. Soh, F.~de~Juan, M.~G. Vergniory, N.~B.~M. Schröter, M.~C. Rahn, D.~Y.
  Yan, J.~Jiang, M.~Bristow, P.~A. Reiss, J.~N. Blandy, Y.~F. Guo, Y.~G. Shi,
  T.~K. Kim, A.~McCollam, S.~H. Simon, Y.~Chen, A.~I. Coldea, and A.~T.
  Boothroyd.
\newblock Ideal {Weyl} semimetal induced by magnetic exchange.
\newblock {\em Physical Review B}, 100(20):201102, nov 2019.

\bibitem{Mewis1978}
A.~Mewis.
\newblock {ABX}-{Verbindungen} mit {Ni}$_2${In}-{Struktur}. {Darstellung} und
  {Struktur} der {Verbindungen} {CaCuP(As)}, {SrCuP(As)}, {SrAgP(As)} und
  {EuCuAs}.
\newblock {\em Z. Naturforsch.}, 33B:983--986, 1978.

\bibitem{Tomuschat1984}
C.~Tomuschat and H.-U. Schuster.
\newblock Magnetische {Eigenschaften} der {Verbindungsreihe} {EuBX} mit {B} =
  {Element} der ersten {Neben}- und {X} = {Element} der f\"{u}nften
  {Hauptgruppe}.
\newblock {\em Z. anorg. allg. Chem.}, 518:161--167, 1984.

\bibitem{du2015dirac}
Y.~Du, B.~Wan, D.~Wang, L.~Sheng, C.-G. Duan, and X.~Wan.
\newblock Dirac and {Weyl} semimetal in \textit{X}\textit{Y}{Bi} (\textit{X} =
  {Ba}, {Eu}; \textit{Y} = {Cu}, {Ag} and {Au}).
\newblock {\em Sci. Rep.}, 5:14423, 2015.

\bibitem{tong2014magnetic}
J.~Tong, J.~Parry, Q.~Tao, G.-H. Cao, Z.-A. Xu, and H.~Zeng.
\newblock Magnetic properties of {EuCuAs} single crystal.
\newblock {\em J. Alloys and Compounds}, 602:26--31, 2014.

\bibitem{nakamura_thermoelectric_2023}
Naoto Nakamura, Yosuke Goto, Yuki Nakahira, Akira Miura, Chikako Moriyoshi,
  Chul-Ho Lee, Hidetomo Usui, and Yoshikazu Mizuguchi.
\newblock Thermoelectric {Properties} of {Zintl} {Arsenide} {EuCuAs}.
\newblock {\em J. Electron. Mater.}, 52:3121--3131, feb 2023.

\bibitem{EuAgAs_DFT_Magnetization}
Antu Laha, Ratnadwip Singha, Sougata Mardanya, Bahadur Singh, Amit Agarwal,
  Prabhat Mandal, and Z.~Hossain.
\newblock Topological hall effect in the antiferromagnetic dirac semimetal
  {EuAgAs}.
\newblock {\em Phys. Rev. B}, 103:L241112, Jun 2021.

\bibitem{Jin_EuAgAs_DFT}
Yahui Jin, Xu-Tao Zeng, Xiaolong Feng, Xin Du, Weikang Wu, Xian-Lei Sheng,
  Zhi-Ming Yu, Ziming Zhu, and Shengyuan~A. Yang.
\newblock Multiple magnetism-controlled topological states in {EuAgAs}.
\newblock {\em Phys. Rev. B}, 104:165424, Oct 2021.

\bibitem{EuAuAs_DFT_Magnetization}
S.~Malick, J.~Singh, A.~Laha, V.~Kanchana, Z.~Hossain, and D.~Kaczorowski.
\newblock Electronic structure and physical properties of {EuAuAs} single
  crystal.
\newblock {\em Phys. Rev. B}, 105:045103, Jan 2022.

\bibitem{EuCuP_Expt}
Jing Wang, Jianlei Shen, Yibo Wang, Tingting Liang, Xiaoyu Wang, Ruiqi Zu, Shen
  Zhang, Qingqi Zeng, Enke Liu, and Xiaohong Xu.
\newblock Anisotropic magneto-transport behavior in a hexagonal ferromagnetic
  {EuCuP} single crystal.
\newblock {\em J. Alloys and Compounds}, 947:169620, 2023.

\bibitem{EuCuBi_DFT}
Xuhui Wang, Boxuan Li, Liqin Zhou, Long Chen, Yulong Wang, Yaling Yang, Ying
  Zhou, Ke~Liao, Hongming Weng, and Gang Wang.
\newblock Structure, physical properties, and magnetically tunable topological
  phases in topological semimetal {EuCuBi}, 2023.

\bibitem{Boothroyd_book}
Andrew~T. Boothroyd.
\newblock {\em {Principles of neutron scattering from condensed matter}}.
\newblock Oxford University Press, 2020.

\bibitem{Yoshimori1959}
A.~Yoshimori.
\newblock A new type of antiferromagnetic structure in the rutile type crystal.
\newblock {\em J. Phys. Soc. Jpn}, 14:807--821, Jun 1959.

\bibitem{Nagamiya1962}
T.~Nagamiya, K.~Nagata, and Y.~Kitano.
\newblock Magnetization process of a screw spin system.
\newblock {\em Prog. Theor. Phys.}, 27:1253--1271, Jun 1962.

\bibitem{Kitano1964}
Y.~Kitano and T.~Nagamiya.
\newblock Magnetization process of a screw spin system. {II}.
\newblock {\em Prog. Theor. Phys.}, 31:1--43, Jan 1964.

\bibitem{Nagamiya1967}
T.~Nagamiya.
\newblock Helical spin ordering --- 1 \textsc{T}heory of helical spin
  configurations.
\newblock {\em Solid State Physics}, 20:305--411, 1967.

\bibitem{Robinson1970}
J.~M. Robinson and P.~Erd\"{o}s.
\newblock Behavior of helical spin structures in applied magnetic fields.
\newblock {\em Phys. Rev. B}, 2:2642--2648, Oct 1970.

\bibitem{Johnston2017}
D.~C. Johnston.
\newblock Magnetic structure and magnetization of helical antiferromagnets in
  high magnetic fields perpendicular to the helix axis at zero temperature.
\newblock {\em Phys. Rev. B}, 96:104405, Sep 2017.

\bibitem{Koshelev2022}
A.~E. Koshelev.
\newblock Phenomenological theory of the {90$^\circ$} helical state.
\newblock {\em Phys. Rev. B}, 105:094441, Mar 2022.

\bibitem{Rahn_2018_ECA}
M.~C. Rahn, J.-R. Soh, S.~Francoual, L.~S.~I. Veiga, J.~Strempfer, J.~Mardegan,
  D.~Y. Yan, Y.~F. Guo, Y.~G. Shi, and A.~T. Boothroyd.
\newblock Coupling of magnetic order and charge transport in the candidate
  dirac semimetal {EuCd$_2$As$_2$}.
\newblock {\em Phys. Rev. B}, 97:214422, Jun 2018.

\bibitem{Soh_2019_ECS}
J.-R. Soh, C.~Donnerer, K.~M. Hughes, E.~Schierle, E.~Weschke, D.~Prabhakaran,
  and A.~T. Boothroyd.
\newblock Magnetic and electronic structure of the layered rare-earth pnictide
  {EuCd$_2$Sb$_2$}.
\newblock {\em Phys. Rev. B}, 98:064419, Aug 2018.

\bibitem{Blawat_2022_EZA}
Joanna Blawat, Madalynn Marshall, John Singleton, Erxi Feng, Huibo Cao, Weiwei
  Xie, and Rongying Jin.
\newblock Unusual electrical and magnetic properties in layered
  {EuZn$_2$As$_2$}.
\newblock {\em Adv. Quantum Technol.}, 5(6):2200012, 2022.

\bibitem{Gui_2019_ESP}
Xin Gui, Ivo Pletikosic, Huibo Cao, Hung-Ju Tien, Xitong Xu, Ruidan Zhong,
  Guangqiang Wang, Tay-Rong Chang, Shuang Jia, Tonica Valla, Weiwei Xie, and
  Robert~J. Cava.
\newblock A new magnetic topological quantum material candidate by design.
\newblock {\em ACS Cent. Sci.}, 5(5):900--910, 2019.
\newblock PMID: 31139726.

\bibitem{Marshall_2021_EMB}
Madalynn Marshall, Ivo Pletikosić, Mohammad Yahyavi, Hung-Ju Tien, Tay-Rong
  Chang, Huibo Cao, and Weiwei Xie.
\newblock Magnetic and electronic structures of antiferromagnetic topological
  material candidate {EuMg$_2$Bi$_2$}.
\newblock {\em J. Appl. Phys.}, 129(3):035106, 2021.

\bibitem{Pakhira_2022_EMS}
Santanu Pakhira, Farhan Islam, Evan O'Leary, M.~A. Tanatar, Thomas Heitmann,
  Lin-Lin Wang, R.~Prozorov, Adam Kaminski, David Vaknin, and D.~C. Johnston.
\newblock A-type antiferromagnetic order in semiconducting {EuMg$_2$Sb$_2$}
  single crystals.
\newblock {\em Phys. Rev. B}, 106:024418, Jul 2022.

\bibitem{Riberolles_2021_EIA}
S.~X.~M. Riberolles, T.~V. Trevisan, B.~Kuthanazhi, T.~W. Heitmann, F.~Ye,
  D.~C. Johnston, S.~L. Bud’ko, D.~H. Ryan, P.~C. Canfield, A.~Kreyssig,
  A.~Vishwanath, R.~J. McQueeney, L.~L. Wang, P.~P. Orth, and B.~G. Ueland.
\newblock Magnetic crystalline-symmetry-protected axion electrodynamics and
  field-tunable unpinned {Dirac} cones in {EuIn$_2$As$_2$}.
\newblock {\em Nat. Commun.}, 12(1):999, February 2021.

\bibitem{Soh_2023_EIA}
Jian-Rui Soh, Alessandro Bombardi, Fr\'{e}d\'{e}ric Mila, Marein~C. Rahn,
  Dharmalingam Prabhakaran, Sonia Francoual, Henrik~M. R{\o}nnow, and Andrew~T.
  Boothroyd.
\newblock Understanding unconventional magnetic order in a candidate axion
  insulator by resonant elastic x-ray scattering.
\newblock {\em Nat. Commun.}, XX(X):xxx, 2023.

\bibitem{Sangeetha_2019_ENA}
N.~S. Sangeetha, V.~Smetana, A.-V. Mudring, and D.~C. Johnston.
\newblock Helical antiferromagnetic ordering in {EuNi$_{1.95}$As$_2$} single
  crystals.
\newblock {\em Phys. Rev. B}, 100:094438, Sep 2019.

\bibitem{Iida_2019_ERFA}
K.~Iida, Y.~Nagai, S.~Ishida, M.~Ishikado, N.~Murai, A.~D. Christianson,
  H.~Yoshida, Y.~Inamura, H.~Nakamura, A.~Nakao, K.~Munakata, D.~Kagerbauer,
  M.~Eisterer, K.~Kawashima, Y.~Yoshida, H.~Eisaki, and A.~Iyo.
\newblock Coexisting spin resonance and long-range magnetic order of {Eu} in
  {EuRbFe$_4$As$_4$}.
\newblock {\em Phys. Rev. B}, 100:014506, Jul 2019.

\bibitem{Gaudet_2021_NdAlSi}
Jonathan Gaudet, Hung-Yu Yang, Santu Baidya, Baozhu Lu, Guangyong Xu, Yang
  Zhao, Jose~A. Rodriguez-Rivera, Christina~M. Hoffmann, David~E. Graf,
  Darius~H. Torchinsky, Predrag Nikolić, David Vanderbilt, Fazel Tafti, and
  Collin~L. Broholm.
\newblock Weyl-mediated helical magnetism in {NdAlSi}.
\newblock {\em Nat. Mater.}, 20(12):1650--1656, December 2021.

\bibitem{fullprof}
J.~Rodr\'{i}guez-Carvajal.
\newblock {\em Physica B}, 192:55--69, 1993.
\newblock http://www.ill.eu/site/fullprof/.

\bibitem{QureshiMag2Pol}
N.~Qureshi.
\newblock {{\it Mag2Pol}: a program for the analysis of spherical neutron
  polarimetry, flipping ratio and integrated intensity data}.
\newblock {\em J. Appl. Crystallogr.}, 52(1):175--185, Feb 2019.

\bibitem{Lelievre2005}
E.~Leli\'{e}vre-Berna, E.~Bourgeat-Lami, P.~Fouilloux, B.~Geffray, Y.~Gibert,
  K.~Kakurai, N.~Kernavanois, B.~Longuet, F.~Mantegezza, M.~Nakamura, S.~Pujol,
  L.-P. Regnault, F.~Tasset, M.~Takeda, M.~Thomas, and X.~Tonon.
\newblock Advances in spherical neutron polarimetry with cryopad.
\newblock {\em Physica B: Condens. Matter}, 356(1):131--135, 2005.

\bibitem{VASP1}
G.~Kresse and J.~Furthm\"uller.
\newblock Efficient iterative schemes for ab initio total-energy calculations
  using a plane-wave basis set.
\newblock {\em Phys. Rev. B}, 54:11169--11186, Oct 1996.

\bibitem{VASP2}
G.~Kresse and J.~Furthmüller.
\newblock Efficiency of ab-initio total energy calculations for metals and
  semiconductors using a plane-wave basis set.
\newblock {\em Computational Materials Science}, 6(1):15--50, 1996.

\bibitem{Perdew96}
John~P. Perdew, Kieron Burke, and Matthias Ernzerhof.
\newblock Generalized gradient approximation made simple.
\newblock {\em Phys. Rev. Lett.}, 77:3865--3868, Oct 1996.

\bibitem{PhysRevB.13.5188}
H.~J. Monkhorst and J.~D. Pack.
\newblock {\em Phys. Rev. B}, 13:5188--5192, Jun 1976.

\bibitem{Yosida-Miwa-JAP}
K.~Yosida and H.~Miwa.
\newblock Magnetic ordering in the ferromagnetic rare-earth metals.
\newblock {\em J. Appl. Phys.}, 32:S8--S12, Mar 1961.

\bibitem{NDD9}
J.-R. Soh, J.~A. Rodriguez-Velamazan, A.~Stunault, and A.T. Boothroyd.
\newblock Structure of a spin-flop phase in the {Weyl} semimetal {EuCuAs},
  2020.
\newblock 10.5291/ILL-DATA.5-41-1048.

\bibitem{SNPD3}
J.-R. Soh, J.~A. Rodriguez-Velamazan, A.~Stunault, and A.T. Boothroyd.
\newblock Is the magnetic structure of {EuCuAs} a transverse helix or a
  collinear antiferromagnet?, 2020.
\newblock 10.5291/ILL-DATA.5-54-368.

\bibitem{ISIS_EuCuAs}
J.-R. Soh, D.~Prabhakaran, P.~Manuel, and A.T. Boothroyd.
\newblock Ground state magnetic structure of {EuCuAs}, 2019.
\newblock 10.5286/ISIS.E.RB1820237.

\bibitem{LET_EuCuAs}
A.~T. Boothroyd.
\newblock Spin dynamics in the magnetic {Weyl} semimetal {EuCuAs}, 2021.
\newblock 10.5286/ISIS.E.RB2090057.

\bibitem{Merlin_EuCuAs}
A.~T. Boothroyd, J-R. Soh, J.~Sun, and D.~Prabhakaran.
\newblock Spin excitations in the candidate {Weyl} semimetal {EuCuAs}, 2019.
\newblock 10.5286/ISIS.E.RB1920514-1.

\end{thebibliography}

\section{Acknowledgments}
\begin{acknowledgments}
The authors  wish  to  thank Gareth Nisbet, Robert Pocock, Dan Porter, Sid Parameswaran, Steve Simon and Ross Stewart for discussions, Toby Perring for providing the software used to  powder-average the spin-wave spectrum presented in Fig.~\ref{fig:Figure_5}\textbf{b}, and  Pascal Manuel, Dmitry Khalyavin and Fabio Orlandi for help with the powder diffraction experiment on WISH at the ISIS Facility. The proposal numbers for the data presented in this manuscript are 5-41-1048 (D9, ILL~\cite{NDD9}), 5-54-368 (D3, ILL~\cite{SNPD3}), 20220501 (SIS-ULTRA, SLS), RB1820237 (WISH, ISIS~\cite{ISIS_EuCuAs}), RB2090057 (LET, ISIS~\cite{LET_EuCuAs}) RB1920514 (Merlin, ISIS~\cite{Merlin_EuCuAs}), MT20347-1 (I16, DLS). D.P. and A.T.B. acknowledge support from the Oxford–ShanghaiTech collaboration project. This work was supported by the U.K. Engineering and Physical Sciences Research Council, grant no. EP/M020517/1. J.-R.S. acknowledges support from the Singapore National Science Scholarship, Agency for Science Technology and Research and the European Research Council (HERO, Grant No. 810451). 
\end{acknowledgments}

\section{Author contributions}
J.-R.S. and A.T.B. conceived the experiments. D.P. grew the single crystals, and J.-R.S., I.V. and D.P. characterised and performed bulk measurements on the crystals. Unpolarized neutron powder and single crystal diffraction was carried out by J.-R.S., J.S., A.T.B., J.A.R.-V. and O.F. The REXS experiment was conducted by J.-R.S., J.S. and A.B., and J.-R.S. and X.P. performed the ARPES experiment.  The INS measurements were made by C.B., M.D.L., H.C.W. and A.T.B., and  the SNP experiment was performed by J.-R.S., J.A.R.-V. and A.S..  The \textit{ab initio} electronic structure calculations and interpretation were performed by I.S.-R., F.d.J. and M.G.V., and A.T.B.  performed the mean-field and spin-wave analysis.  All authors reviewed the manuscript.

\section{Competing interests}
The authors declare no competing interests.

\end{document}